\documentclass[a4paper,fleqn,usenatbib]{mnras}

\usepackage{newtxtext,newtxmath}

\usepackage[T1]{fontenc}
\usepackage{ae,aecompl}

\usepackage{graphicx}	
\usepackage{amsmath}	
\usepackage{amssymb}	
\usepackage{xspace}        

\newcommand{\vlbipsr}{PSR~J0437$-$4715\xspace}

\title[Two AGNs near PSR J0437$-$4715]{Revealing two radio active galactic nuclei extremely near PSR J0437$-$4715}

\author[Z. Li et al.]{Zhixuan Li,$^{1,7,8}$
Jun Yang,$^{2,1,3}$\thanks{E-mail: jun.yang@chalmers.se}
Tao An,$^{3,9}$\thanks{E-mail: antao@shao.ac.cn}
Zsolt Paragi,$^{4}$
Adam Deller,$^{5}$ \newauthor
Cormac Reynolds,$^6$
Xiaoyu Hong,$^{3,9}$
Jiancheng Wang,$^{1,8}$
Hao Ding,$^{3,7,9}$
Bo Xia,$^{3,9}$ \newauthor
Zhen Yan$^{3,9}$ 
and Li Guo$^{3,9}$
\\
\\
$^{1}$Yunnan Observatories, Chinese Academy of Sciences, 650216 Kunming, Yunnan, P.R. China \\
$^{2}$Department of Space, Earth and Environment, Chalmers University of Technology, Onsala Space Observatory \\
SE-439 92 Onsala, Sweden \\
$^{3}$Shanghai Astronomical Observatory, Chinese Academy of Sciences, 200030 Shanghai, P.R. China\\
$^{4}$Joint Institute for VLBI ERIC (JIVE), Postbus 2, NL-7990 AA Dwingeloo, the Netherlands \\
$^{5}$Centre for Astrophysics \& Supercomputing, Swinburne University of Technology, John St, Hawthorn VIC 3122, Australia \\
$^{6}$CSIRO Astronomy and Space Science, Kensington, WA 6151, Australia \\
$^{7}$University of Chinese Academy of Sciences, 19A Yuquan Road, Shijingshan District, 100049 Beijing, P.R. China \\
$^{8}$Key Laboratory for the Structure and Evolution of Celestial Objects, Chinese Academy of Sciences, 650216 Kunming, P.R. China \\
$^{9}$Key Laboratory of Radio Astronomy, Chinese Academy of Sciences, 210009 Nanjing, P.R. China \\
}
\date{Accepted 2018 XXX. Received 2018 YYY; in original form 2018 ZZZ}

\pubyear{2018}

\begin{document}
\label{firstpage}
\pagerange{\pageref{firstpage}--\pageref{lastpage}}
\maketitle

\begin{abstract}
Newton's gravitational constant $G$ may vary with time at an extremely low level. The time variability of $G$ will affect the orbital motion of a millisecond pulsar in a binary system and cause a tiny difference between the orbital period-dependent measurement of the kinematic distance and the direct measurement of the annual parallax distance. \vlbipsr is the nearest millisecond pulsar and the brightest at radio. To explore the feasibility of achieving a parallax distance accuracy of one light-year, comparable to the recent timing result, with the technique of differential astrometry, we searched for compact radio sources quite close to \vlbipsr. Using existing data from the Very Large Array and the Australia Telescope Compact Array, we detected two sources with flat spectra, relatively stable flux densities of 0.9 and 1.0~mJy at 8.4~GHz and separations of 13 and 45~arcsec. With a network consisting of the Long Baseline Array and the Kunming 40-m radio telescope, we found that both sources have a point-like structure and a brightness temperature of $\geq$10$^7$~K. According to these radio inputs and the absence of counterparts in the other bands, we argue that they are most likely the compact radio cores of extragalactic active galactic nuclei rather than Galactic radio stars. The finding of these two radio active galactic nuclei will enable us to achieve a sub-pc distance accuracy with the in-beam phase-referencing very-long-baseline interferometric observations and provide one of the most stringent constraints on the time variability of $G$ in the near future.    
\end{abstract}




\section{Introduction}
\label{sec1}

Millisecond pulsars (MSPs) are neutron stars that have been `recycled' via accretion of material from a companion \citep{Alpar1982} and that spin with a period usually shorter than about ten milliseconds. Compared to the normal pulsars that spin more slowly, MSPs have much more stable rotation. Thus, their radio pulses can be used as extremely accurate clocks for the study of relativistic gravitation \citep[e.g.,][]{Manchester2015}. 

\vlbipsr is an MSP discovered by \citet{Johnston1993} with the 64-m Parkes radio telescope in Australia. \vlbipsr is the nearest MSP, and the brightest in the radio band. It has a statistical mean flux density of 150~mJy at 1.4~GHz \citep[e.g.,][]{Dai2015}. As one of the most stable and precisely timed pulsars, it is a cornerstone of efforts to use a pulsar timing array to search for nanohertz-frequency gravitational waves \citep[e.g.,][]{Shannon2015}, and has been intensively observed with the Parkes telescope for more than 20 yr. Its timing model has been continuously improved by timing analyses \citep[e.g.,][]{Verbiest2008}.  By using the longer time baseline data and modelling the non-stationary noise, \citet{Reardon2016} significantly improved the timing measurements of the proper motion $\mu$, the orbital period $P_\mathrm{b}$ and the orbital period derivative $\dot{P}_\mathrm{b}$. The observed orbital period derivative is dominated by a `kinematic' term due to the transverse motion of the pulsar; accordingly, the timing observables can be used to obtain a kinematic distance of $D_\mathrm{k}=\frac{c}{\mu^2}\frac{\dot{P_\mathrm{b}}}{P_\mathrm{b}}=156.79\pm0.25$~pc. To date, this is the most precise distance measurement to a radio pulsar. 

The distance to a radio pulsar can also be accurately measured with multi-epoch very-long-baseline interferometric (VLBI) observations of the annual parallax $\pi$. The position measurements are made with respect to a nearby extra-galactic source whose position is assumed to be stationary, that is, a phase-referencing calibrator. The differential VLBI astrometry on \vlbipsr can provide not only an accurate parallax distance $D_\mathrm{\pi}$ but also a stringent constraint on the stability of Newton's gravitational constant $G$, i.e. $\dot{G}$. This is because the difference between $D_\mathrm{\pi}$ and $D_\mathrm{k}$ represents the maximum contribution from the potential $\dot{G}$. \citet{Deller2008} measured $D_\pi = \frac{1}{\pi}=156.3\pm1.3$~pc ($\pi=6396\pm54$~$\umu$as) for \vlbipsr with the Australia Long Baseline Array (LBA) astrometry at 8.4~GHz and estimated a tight upper limit of  $\dot{G}/G=(-5\pm26)\times10^{-13}$~yr$^{-1}$  together with the earlier measurement of $D_\mathrm{k}=157.0\pm2.4$~pc \citep{Verbiest2008}. Because the $D_\mathrm{k}$ has been recently measured with unprecedented precision, 0.25 pc \citep{Reardon2016}, it is becoming a key question in the study of the $\dot{G}$ whether it is feasible to also gain a much more accurate $D_\pi$ in the near future. 

The easiest and most reliable way to improve the VLBI parallax accuracy is to use a phase-referencing source as close to the target source as possible \citep[e.g., the PSR$\pi$ project, ][]{Deller2016}. The small angular separation means that differential propagation effects through the ionosphere and troposphere are minimised, and moreover ensures that all the radio telescopes can observe both sources simultaneously. Thus, systematic phase errors affecting the target can be calibrated and removed very accurately via solutions obtained from the phase-referencing source. Based on the in-beam phase-referencing technique, a parallax accuracy of $\sim$15~$\umu$as was achieved by \citep{Deller2013} for PSR J2222$-$0137. According to the known pulsar parallax measurements, the best attainable parallax accuracy has an empirical dependence of about 1--2~$\umu$as per arc-minute on the calibrator-target separation \citep{Chatterjee2004, Deller2013, Kirsten2015}. However, this ``floor" is the best attainable result of a perfect calibrator: one that is bright enough to enable calibration solutions on short integration time, and that does not exhibit any significant source structure variations.

This paper is organised in the following sequence. Section \ref{sec2} describes the radio experiments we used and the data reduction steps we followed. Section \ref{sec3} reports the imaging and data analysis results. Section \ref{sec4} discusses the identification of two radio sources that we detected in the radio observations, the feasibility of achieving a sub-pc parallax precision on \vlbipsr with in-beam phase-referencing VLBI observations and the implications for a more stringent constraint on $\dot{G}$.  Section \ref{sec5}  gives the final conclusions.

\section{Observations and data reduction}
\label{sec2}

\begin{table*}    
\caption{\label{tab1}Summary of the radio observations of \vlbipsr field. }
\begin{center}
\begin{tabular}{*9{c}}
\hline\hline
Project   & Array         & $N_\mathrm{ant}$   & Date                &  Freq.            & Bw               & Time               &  Beam                                                                &   Sensitivity       \\
Code      &                  &                                 &                         &  (GHz)           & (MHz)          &  (min)             &  FWHM                                                               &   (mJy beam$^{-1}$)    \\
\hline%
AH0595   & VLA         & 12                           & 1996 Oct 12     &  1.435             & 100              & 76                  & $5\farcs40\times0\farcs79$ at $-0\fdg22$           & 0.11                 \\
AH0595   & VLA         & 15                           & 1996 Oct 12     &  4.860             & 100              & 100                & $6\farcs40\times0\farcs95$ at $+0\fdg30$          & 0.05                \\
\hline
V190A     & ATCA       & 6                           & 2006 May 13     & 8.428              &40               & 596                 & $2\farcs28\times 1\farcs36$ at $-0\fdg47$    &0.10 \\
V190E     & ATCA       & 5                           & 2006 Nov 16     & 8.428             &40                 & 688                 & $6\farcs05\times 4\farcs03$ at $-2\fdg74$        &0.18 \\
V190G    & ATCA        & 5                          & 2007 Mar 22     & 8.428             &40                  & 694                 & $10\farcs90\times 7\farcs39$ at $+0\fdg41$       &0.13 \\
V190K    & ATCA        & 6                           & 2007 Nov 12.    & 8.428            &40                & 697                   & $2\farcs09\times 1\farcs44$ at $-0\fdg22$     &0.09 \\
V190M   & ATCA       & 6                             & 2009 Dec 12    & 8.457            &384                 & 579                & $2\farcs33\times 1\farcs35$ at $-5\fdg86$      &0.14 \\
V190O   & ATCA        & 5                           & 2010 Oct 27      & 8.424              &50              & 609                   &  $27\farcs1\times 20\farcs4$ at $-81\fdg70$      &0.13 \\
\hline
V539      & LBA$+$KM   & 5                             &  2015 Nov 18   &  2.285             &   32              &  208               & $0\farcs015\times0\farcs0056$ at $+41\fdg3$  & 0.09                \\
\hline\hline
\end{tabular}
\end{center}
\flushleft
Note. Columns give (1) project code, (2) array name, (3) total number of telescopes ($N_\mathrm{ant}$), (4) observation date, (5) observation frequency in GHz, (6) total bandwidth in MHz, (7) total on-source time in min, (8) beam FWHM (full width to half magnitude) and (9) image sensitivity in mJy~beam$^{-1}$. 
\end{table*}

To search for extragalactic compact radio sources quite close to \vlbipsr, we re-analysed an early Very Large Array (VLA) experiment (project code: AH0595) provided by the NRAO Science Data Archive\footnote{URL: https://archive.nrao.edu/archive/advquery.jsp}.  After finding two candidates within one arcmin circle from \vlbipsr, we performed in-beam phase-referencing VLBI observations of them with the Australia Long Baseline Array (LBA) plus the Chinese Kunming 40-m radio telescope at 2.3~GHz (project code: V539).  Both the VLA data and the VLBI data were calibrated in \textsc{aips} \citep[Astronomical Image Processing System,][]{Greisen2003}. The iterative loop of source deconvolution and self-calibration was conducted in \textsc{difmap} \citep{Shepherd1994}. To study their flux density stability, we also reduced the existing interferometric data (project code: V190) that comprised a multi-epoch series spanning about 4.5 yr and as observed by the Australia Telescope Compact Array (ATCA) at 8.4~GHz during the LBA observations of \vlbipsr \citep{Deller2008}. The ATCA data were calibrated in the software package of \textsc{miriad} \citep{Sault1995}. 

The observational information about these experiments and the final intensity image parameters are listed in Table~\ref{tab1}. The total flux densities of \vlbipsr and the two faint sources, named R1 and R2, and their observation dates and frequencies are reported in Table~\ref{tab2}.   

\subsection{VLA data at 1.4 and 5 GHz}
\label{sec2-1}

The VLA experiment is one of the deepest observations of \vlbipsr. It was performed in the array configuration of A and D on 1996 October 12. The whole array was divided into two sub-arrays to observe \vlbipsr simultaneously at both 1.4 and 5~GHz. Because of the quite low declination of \vlbipsr, both sub-arrays had relatively high resolution (up to sub-arcsec) only in the East-West direction. The source J0440$-$4333 (B0440$-$435) was observed as the phase-referencing calibrator. The multi-frequency monitoring observations performed by \citet{Tingay2003} with the ATCA between 1996 October and 2000 February show that J0440$-$4333 was quite stable with a variability index of 0.02 at 1.38~GHz and 0.10 at 4.80~GHz. We also used J0440$-$4333 as the flux density calibrator in the subsequent data reduction, given its known ATCA mean flux density measurements of 4.53~Jy at 1.38~GHz and 3.19~Jy at 4.80~GHz. According to the ATCA monitoring results, we took $\sim$2 per cent as 1$\sigma$ systematic flux density error at both bands. Following the standard VLA data reduction recipe in the \textsc{aips} cookbook\footnote{http://www.aips.nrao.edu/cook.html}, we removed some bad data at the beginning of each scan, set the flux density of J0440$-$4333, ran both self-calibration and bandpass calibration on the calibrator, and applied all the solutions to both the calibrator and the pulsar. 

In the final imaging process of \vlbipsr, we excluded the data on the short baselines of $\leq$20 kilo-wavelengths at 1.4~GHz to avoid some imaging errors caused by far-field bright sources. The self-calibrations were applied first with the phase only and then both the phase and the amplitude. The calibration helped to reduce the image noise level by a factor of about four at 5~GHz. In the residual map of 1.4~GHz, the calibration also allowed us to remove some significant noise peaks (up to 19~mJy~beam$^{-1}$). 

\subsection{New VLBI experiment at 2.3~GHz}
\label{sec2-2}

We performed in-beam phase-referencing VLBI observations of the two candidate sources at 2.3~GHz with the Australia LBA and the Chinese Kunming 40-m radio telescope \citep{Hao2010} on 2015 November 18. This was the first time that the Kunming telescope had carried out an astronomical VLBI experiment together with the VLBI network of the southern hemisphere. The LBA participating stations included the phased-up array of the ATCA, the Parkes 64-m radio telescope, the Hobart 26-m radio telescope, and the Ceduna 30-m radio telescope. 

The VLBI observations lasted about six hours. The synthesised beam of the ATCA had a full width at half maximum  (FWHM) of $13\farcs3\times4\farcs8$, which is quite narrow and insufficient to encompass \vlbipsr and sources R1 and R2 simultaneously. In view of this limitation, we provided a separate schedule for the ATCA correlator to cycle its correlation phase centre among the three sources with an interval of two minutes. The remaining telescopes have quite wide beams and thus observed all the three sources with the same pointing centre. The observations had dual polarisation and an observation bandwidth of 32~MHz per polarisation.  

The data were correlated in three passes with Distributed FX correlator \citep[DiFX;][]{Deller2007}. A pulsar gate, using the publicly available ephemeris for \vlbipsr, was used for the pass at the pulsar position. The correlator output data had one second integration time per data point, four sub-bands per polarisation and 32 frequency points per sub-band. 

The a-priori amplitude calibration was done by three steps. Using the attached auto-correlation data, the cross-correlation amplitude was properly scaled and corrected. According to the nominal values of system equivalent flux density provided by each station, the amplitude gain solutions were derived and applied.  

Because \vlbipsr is a perfect point source for both the ATCA and the VLBI network, we also used it as a flux density calibrator to improve the VLBI amplitude calibration. As an independent array, the ATCA has also provided the simultaneous interferometric data. We reduced the zoom-in band data (64-MHz bandwidth) separately in \textsc{miriad} to measure the total flux density of \vlbipsr. The correlation amplitude of the ATCA data were already properly calibrated by dedicated calibrator observations, so only self-calibration was performed for the ATCA data.  

For the LBA data sets, delay and phase calibration was performed via running fringe-fitting on \vlbipsr data with a solution interval of one minute. In our bandpass calibration, only phase solutions were solved and applied. After these calibrations, the final solution tables were copied from the pulsar data set to the other two weak source data sets. Because different data sets had different source ID for the same source, before we applied the solution tables, we used the \textsc{aips} task \textsc{tabed} to adjust the source ID in the calibration table to the appropriate value as determined in the source table.

When we imaged the two faint VLA sources, we removed the data on the baseline between the ATCA to the Parkes telescope because the shortest baseline had some significantly unwanted flux originating from the nearby \vlbipsr and caused some stripes in the dirty image. In the future VLBI observations, the most sensitive baseline may be saved by correlating only the pulse-off data. Moreover, we removed a certain amount of the ATCA data when the ATCA digital beam was off-source for each data set. With naturally weighting, we achieved an image sensitivity of $1\sigma=0.09$~mJy~beam$^{-1}$ in the final \textsc{clean} images. 

The VLBI positions of the three sources are reported in Table~\ref{tab3}. We measured their positions with respect to \vlbipsr. The position of \vlbipsr was calculated from the LBA astrometric model presented by \citet{Deller2008} and was used as the correlation phase centre. The formal position error is $1\sigma$ $\sim$0.6~mas for sources R1 and R2. The absolute position determination of R1 and R2 is currently limited by the uncertainty in the position of \vlbipsr. We measured the offset of these two sources from the assumed position of \vlbipsr, so that any error in the latter's position is transferred. The absolute position uncertainty was obtained using differential astrometry to a single calibrator, as was the case with the earlier observations of \vlbipsr \citep{Deller2008}; it is difficult to quantify, and is more strongly affected by residual calibration errors than time-dependent effects such as parallax and proper motion. We re-processed the data presented in \citet{Deller2008} along with several unpublished epochs, making use of updated station positions for the LBA, and found a reference position for \vlbipsr which differs by 0.7~mas from that presented in \citet{Deller2008}, as well as a revised proper motion that is more consistent with the timing results of \citet{Reardon2016}, differing by less than 2$\sigma$ in each coordinate. The parallax value is not significantly altered. Because our ultimate goal of precisely measuring the parallax and proper motion of \vlbipsr relies only on differential and not absolute positions, the absolute position uncertainty is not a main concern in this research, and we assume here an uncertainty of 0.7~mas and add this in quadrature to the statistical position errors. In our future astrometric observations, we will include some phase-referenced scans from \vlbipsr to J0439$-$4522 that has a well-determined position with an uncertainty of 0.1~mas in the International Celestial Reference Frame (ICRF) to measure the absolute position of \vlbipsr.

\subsection{Multi-epoch ATCA observations at 8.4~GHz}
\label{sec2-3}
\vlbipsr was observed with the ATCA at 8.4~GHz for a total of seven epochs, and the data are available in the Australia Telescope Online Archive. During these 8.4-GHz observations, the ATCA worked mainly as a phased-up array to participate in the LBA astrometric observations of \vlbipsr \citep{Deller2008}. To search for possible flux density variability in sources R1 and R2, we also reduced these additional ATCA data to measure their total flux densities. The quite long on-source time allowed us to achieve an image sensitivity of 0.1--0.2 mJy beam$^{-1}$ and successfully detected sources R1 and R2 in six epochs. Because of quite rapid phase variation likely due to an instrumental issue, we failed to achieve a proper phase calibration in one epoch (V190N). In the subsequent analysis, the epoch was excluded. 

In the data reduction, we used J0439$-$4522 as the secondary calibrator to run the self-calibration and applied its solutions to \vlbipsr. When the ATCA primary flux density calibrator PKS~$1934-638$ was observed, we measured the flux densities of these three sources directly. Amongst the three accurate flux density measurements, \vlbipsr has a mean flux density of 4.01$\pm$0.15~mJy and a variability index of 0.08. The low variability is as expected because the interstellar scintillation is much weaker at the higher observation frequencies \citep[e.g.,][]{Dai2015}. Because \vlbipsr is relatively faint at 8.4~GHz, neither the phase nor the amplitude calibrations were applicable. When PKS~$1934-638$ was not observed in the remaining three epochs (V190E, V190G and V190M), we took \vlbipsr as the primary flux density calibrator and then measured the flux densities of sources R1 and R2. When the zoom-in band data (64-MHz bandwidth) were not available in the experiment V190M, we reduced the wide-band data (512-MHz bandwidth).  

\begin{table}
\caption{\label{tab2}Summary of the total flux density measurements.} 
\begin{center}
\begin{tabular}{ccrc}
\hline\hline
Name                & MJD                & Flux                        & Frequency         \\
                          &                        &  (mJy)                     & (GHz)               \\
\hline%
J0437$-$4715   & 50369.4           & 166.81$\pm$0.11       & 1.435              \\
R1                     & 50369.4           &   0.97$\pm$0.11       & 1.435             \\
R2                     & 50369.4           &   0.77$\pm$0.11       & 1.435              \\
J0437$-$4715   & 50369.4           & 14.90$\pm$0.05      & 4.860              \\
R1                     & 50369.4           &   1.35$\pm$0.05       & 4.860             \\
R2                     & 50369.4           &   1.67$\pm$0.05       & 4.860              \\
\hline
 J0437-4715  & 53869.1 & 4.04$\pm$0.10  & 8.428  \\
 R1          & 53869.1 & 1.06$\pm$0.10  & 8.428  \\
 R2          & 53869.1 & 0.99$\pm$0.10  & 8.428  \\
 J0437-4715  & 54055.6 & 4.01$\pm$0.18  & 8.428  \\
 R1          & 54055.6 & 0.83$\pm$0.18  & 8.428  \\
 R2          & 54055.6 & 0.88$\pm$0.18  & 8.428  \\
 J0437-4715  & 54181.3 & 4.01$\pm$0.13  & 8.428  \\
 R1          & 54181.3 & 0.90$\pm$0.13  & 8.428  \\
 R2          & 54181.3 & 1.09$\pm$0.13  & 8.428  \\
 J0437-4715  & 54416.6 & 3.75$\pm$0.09  & 8.428  \\
 R1          & 54416.6 & 0.88$\pm$0.09  & 8.428  \\
 R2          & 54416.6 & 1.01$\pm$0.09  & 8.428  \\
 J0437-4715  & 55177.5 & 4.01$\pm$0.15  & 8.457  \\
 R1          & 55177.5 & 1.10$\pm$0.15  & 8.457  \\
 R2          & 55177.5 & 1.00$\pm$0.15  & 8.457  \\
 J0437-4715  & 55496.7 & 4.44$\pm$0.13  & 8.424  \\
 R1          & 55496.7 & 0.79$\pm$0.13  & 8.424  \\
 R2          & 55496.7 & 1.23$\pm$0.13  & 8.424  \\ 

\hline
J0437$-$4715   & 57344.6           & 60.32$\pm$0.15      & 2.285              \\ 
R1                     & 57344.6           &   0.83$\pm$0.09       & 2.285              \\ 
R2                     & 57344.6           &   0.66$\pm$0.09       & 2.285              \\ 

\hline\hline
\end{tabular}
\end{center}
\end{table}  

\begin{table}
\caption{\label{tab3}The positions of  \vlbipsr, sources R1 and R2 measured by the VLBI observations on 2015 November 18. }
\begin{center}
\begin{tabular}{*4{c}}
\hline\hline
Source                &   Right Ascension                    & Declination                                          &  1$\sigma$  \\ 
                           &                                                 &                                                             &   (mas)          \\               
\hline%
J0437$-$4715   &  $04^\mathrm{h}37^\mathrm{m}15\fs98936$     
                                                                              &   $-47\degr15\arcmin09\farcs6696$     &  0.7   \\
R1                     &  $04^\mathrm{h}37^\mathrm{m}16\fs20943$
                                                                              &    $-47\degr14\arcmin56\farcs5046$    & 0.9     \\                                
R2                     &  $04^\mathrm{h}37^\mathrm{m}17\fs77577$          
                                                                              &    $-47\degr14\arcmin28\farcs2398$     & 0.9    \\
\hline\hline
\end{tabular}
\end{center}
\end{table}

\section{Detections of two radio sources}
\label{sec3}

\begin{figure*}
	\centering
	\includegraphics[width=2.1\columnwidth]{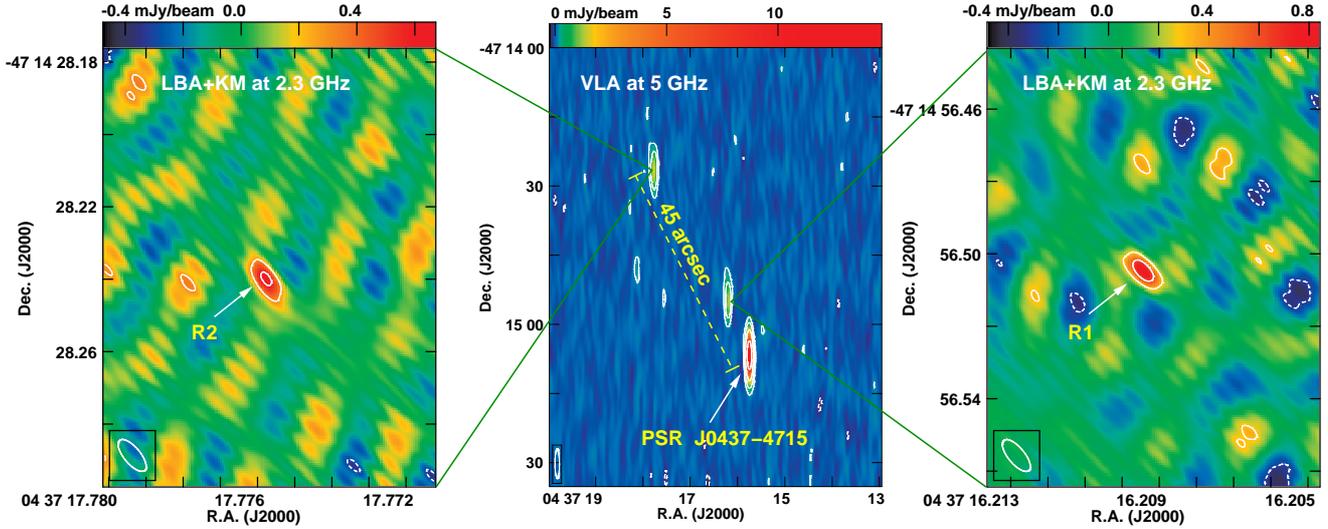}  
    \caption{Radio detection of two compact radio sources (R1 and R2) with a separation of 13 and 45~arcsec apart from the most bright and nearest millisecond pulsar \vlbipsr. Left: LBA$+$KM detection of source R2 at 2.3~GHz. Middle: Wide-field VLA 5-GHz image of the three sources. Right: LBA$+$KM detection of source R1. Contours start from 3$\sigma$ and increase by a factors of -1, 1, 4, 8, ... in the VLA map and -1, 1, 2 in the LBA$+$KM map. The related image parameters are listed in Table~\ref{tab1}. } 
    \label{fig1}
\end{figure*}

The VLA total intensity image at 5~GHz is shown in the middle panel of Fig.~\ref{fig1}. Compared to \vlbipsr, sources R1 and R2 are quite faint. After modelling \vlbipsr with a point source model and applying the self-calibration, the two faint sources are clearly visible features in the residual map. We obtained a signal to noise ratio (SNR) of 29 for R1 and 36 for R2 in the final image with naturally weighting. The angular distance to \vlbipsr is 13 arcsec for R1 and 45 arcsec for R2. Both sources are in a northeast direction and have a difference of $\sim$6 deg in their position angles. At the lower frequency of 1.4~GHz, both sources are fainter, and source R1 is slighter brighter than source R2.  
 
The VLBI images of sources R1 and R2 at 2.3~GHz are displayed in the left and right panels of Fig.~\ref{fig1}. Note that the shortest baseline between the ATCA and the Parkes telescope was excluded to minimise the unwanted signal from \vlbipsr. This helps to improve the image sensitivity by a factor of about two. In the final \textsc{clean} images, we obtained an image noise level of 0.09~mJy beam$^{-1}$ with natural weighting. The image SNRs are 8.6 for source R1 and 6.8 for source R2. 

We consider R1 and R2 as real radio sources as opposed to artefacts from instrumental noise for several reasons. Both sources are the brightest peaks with a SNR $\geq$6 in the VLBI dirty maps which have a size wide enough to cover the 5$\sigma$ error circle of the VLA positions ($5\sigma\la0.5$ arcsec). The image noise patterns surrounding the two peaks resemble the synthesised beam and show some negative peaks which are located at the positions of negative side lobes and include the deepest one. After the deconvolution of the two sources, the remaining noise peaks have a SNR $\leq$4.6 in the residual maps. With respect to the VLA positions (our image origins), the VLBI source R1 lies at an offset of 112.7 mas and R2 at an offset of 88.9 mas, both within the 2$\sigma$ error circle. Their VLBI total flux densities are also consistent with the upper limit ($3\sigma\leq1.2$~mJy) of our simultaneous ATCA observations at 2.3~GHz. At a flux density level of $\geq$0.5~mJy, \citet{HerreraRuiz2017} reports that faint VLA sources retain a detection fraction of $\geq$50 per cent at mas resolutions. Therefore, from a statistical viewpoint, it is not surprising for us to detect both in the VLBI images. 

The VLBI images have a resolution 269 times higher than the VLA image. Both sources are still unresolved. Because of their compact morphology, a simple point source model was used in measuring their total flux densities and positions. The VLA images observed simultaneously at 1.4 and 5~GHz show that the flux densities of both sources increase at the higher frequencies; the inferred slightly inverted spectra are consistent with their compact appearance. The VLBI flux density measurements of the two sources are roughly consistent with the extrapolation of the two VLA measurements from 19~yr earlier.   

We also fitted a circular Gaussian model to the VLBI data to derive an upper limit for their angular sizes $\theta_\mathrm{size}$. We estimated $\theta_\mathrm{size}\leq2.2$~mas for R1 and $\theta_\mathrm{size}\leq4.0$~mas for R2. With the upper limits of $\theta_\mathrm{size}$, the lower limits of their brightness temperature $T_\mathrm{b}$ in K can be estimated according to the following equation \citep[e.g.,][]{Condon1982}.
\begin{equation}
T_\mathrm{b} = 1.22\times10^{9}\frac{S_\mathrm{tot}}{\nu_\mathrm{obs}^2\theta_\mathrm{size}^2}(1+z),
\label{eq1}
\end{equation}
where $S_\mathrm{tot}$ is the total flux density in mJy, $\nu_\mathrm{obs}$ is the observation frequency in GHz, $\theta_\mathrm{size}$ is the FWHM of the circular Gaussian model  in mas, and $z$ is redshift. At $z=0$, our estimation gives $T_\mathrm{b}\geq4\times10^7$~K for R1 and $T_\mathrm{b}\geq1\times10^7$~K for R2. 

\begin{figure*}
\centering
	\includegraphics[width=0.9\textwidth, trim=20 40 40 40, clip]{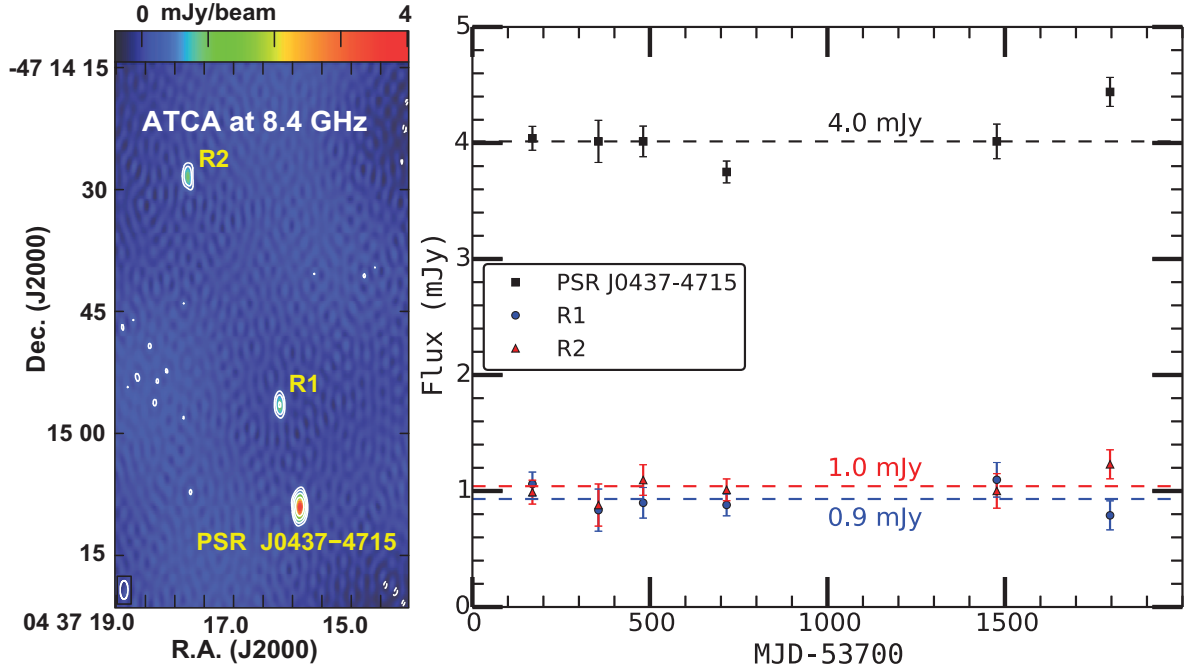}  
    \caption{The radio map and the light curves of \vlbipsr (black), sources R1 (blue) and R2 (red) observed by the ATCA at 8.4~GHz.  In the map observed on 2006 May13,  the contours start from 3$\sigma$ and increase by a factors of -1, 1, 2, 4, ...  The image parameters are listed in Table~\ref{tab1}. The flux density measurements are reported in Table~\ref{tab2}. } 
    \label{fig2}
\end{figure*}

The left panel of Fig.~\ref{fig2} displays the total intensity image of the \vlbipsr field observed by the ATCA in the first 8.4-GHz epoch.  With naturally weighting, the ATCA images have an image sensitivity of 0.1--0.2 mJy beam$^{-1}$. At the VLA positions, both sources are the brightest features with SNRs of 4--10 in the ATCA residual maps after the deconvolution of \vlbipsr.  The 8.4 GHz light curves of sources R1 and R2 are plotted in the right panel of Fig.~\ref{fig2}. Compared to the average flux density, i.e. the blue line for R1 and the red line for R2, no significant deviation is detected over about 4 yr. Together with the agreement between the VLA and the VLBI results, both sources have a relatively stable flux density with a variability index of $\leq$0.15 on time scales of months to years.  

\begin{figure}
\centering
	\includegraphics[width=1.0\columnwidth, trim=20 10 60 50, clip]{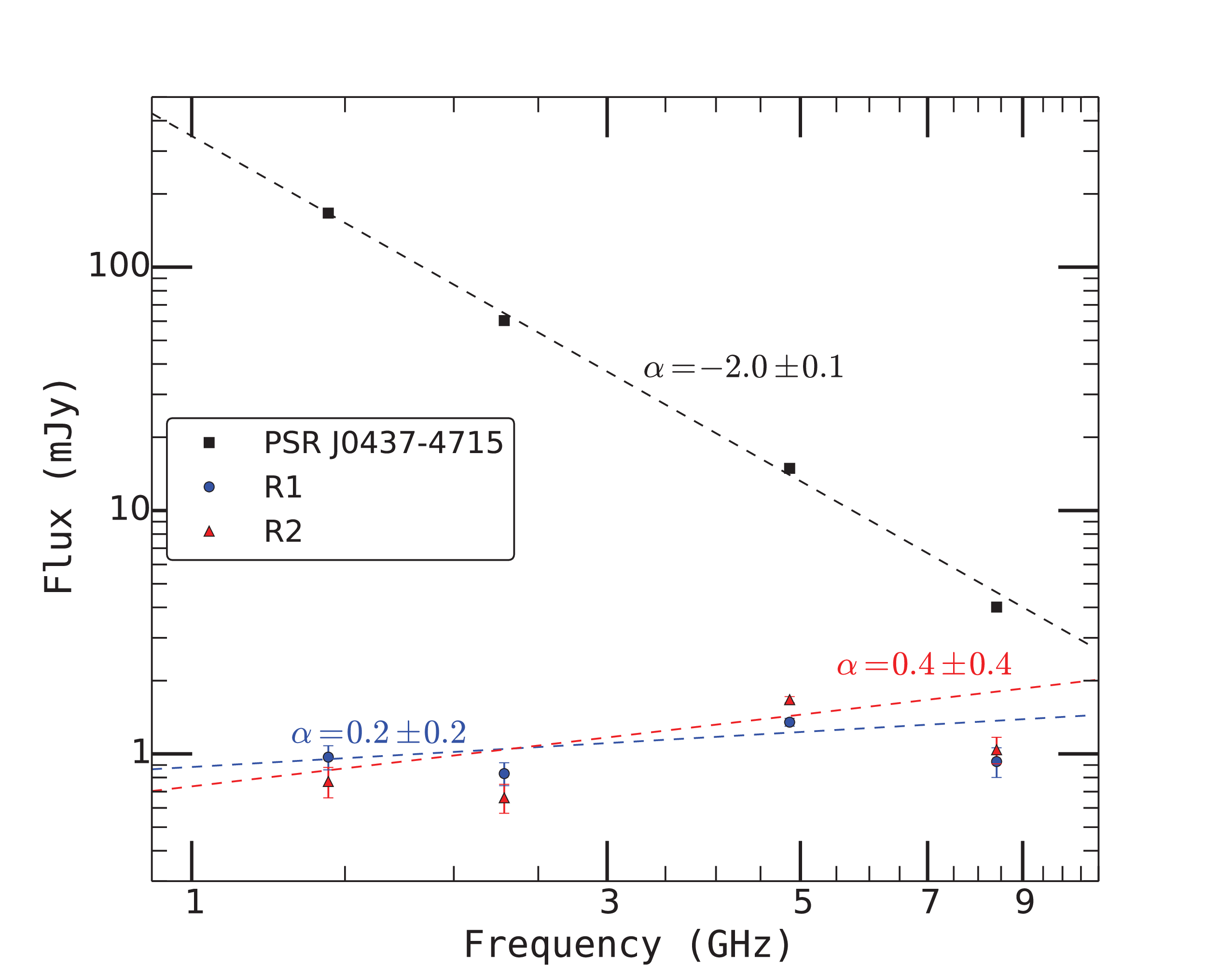}  
    \caption{The radio spectra of sources R1 and R2 observed by the VLA at 1.4 and 5~GHz, the LBA$+$Km at 2.3~GHz and the ATCA at 8.4~GHz. The related flux measurements are summarized in Table~\ref{tab2}. } 
    \label{fig3}
\end{figure}

The multi-frequency flux density measurements are also displayed in Fig.~\ref{fig3}. Assuming that they had a relatively stable luminosity over the 19 yr and a compact source structure, we can estimate their spectral indices ($\alpha$, such that $S_\nu\propto\nu^{\alpha}$). Sources R1 and R2 have quite flat spectra ($\alpha=0.2\pm0.2$ and $\alpha=0.4\pm0.4$  respectively). \vlbipsr has a quite steep spectrum ($\alpha=-2.0\pm0.1$). 

\section{Discussion}
\label{sec4}

\subsection{Identifications: compact radio cores in AGNs}
\label{sec4-1}
The most reasonable identification of sources R1 and R2 is that they are the compact radio cores, i.e. the stationary jet bases, in extragalactic active galactic nuclei (AGNs).  This identification can naturally explain their high brightness temperatures,  compact morphologies, flat radio spectra, and relatively stable luminosities over the 19 yr. Their brightness temperatures are clearly higher than the typical value of $\leq$10$^6$~K observed in thermal emission sources. So, all thermal emission origins can be firmly excluded. Currently, either jets powered by compact objects or shocks formed in a certain astrophysical environments produce relativistic electrons and thus non-thermal synchrotron radio emission. Compared with the shock scenario, the AGN core scenario can naturally explain the point-like morphology and flat spectra of R1 and R2, as cores would generally have a partially self-absorbed spectrum with $\alpha\sim0$. Moreover, shocks are usually associated with non-continuous injection of relativistic electrons and have relatively shorter lives and greater variability; whilst both sources have relatively low variability indices ($\sim$15 per cent)  in the multi-epoch radio observations.  

Sources R1 and R2 cannot be Galactic objects with significantly pulsed radio emission, e.g., pulsars. \vlbipsr has been observed by the Parkes radio telescope for about 20~yr \citep{Reardon2016}. Because both sources are also in the same telescope beam, it is unlikely that the long-term timing observations have missed any pulsed emission. Furthermore, they do not have the very steep radio spectra observed in most pulsars.   

It is also quite difficult to identity sources R1 and R2 as the radio jets powered by stellar-mass compact objects in the Milky Way, such as white dwarfs, neutron stars, and stellar black holes in X-ray binaries. These compact objects usually have luminosities that vary significantly from radio to X-ray bands. We found no X-ray counterparts at the two locations in the image presented by \citet{Zavlin2002} with 18.9~ks High Resolution Camera observations of the \textit{Chandra X-ray Observatory}. \textit{XMM-Newton} MOS1 and MOS2 mosaic images of the field around \vlbipsr in the 0.1--10~keV band, reported by \citet{Bogdanov2013}, also show no counterparts.

Sources R1 and R2 have no infra-red and optical counterparts. This is not an unusual result since the optical counterparts for about two third of radio sources are still missing \citep{Kimball2008}. Neither were found in the all-sky data release of the \textit{Wide-field Infrared Survey Explorer} (\textit{WISE}) on 2012 March 14, with $5\sigma$ flux density upper limits at 3.4, 4.6, 12, and 22~$\umu$m of 0.08, 0.11, 1 and 6~mJy respectively.  We searched for their optical counterparts in the high sensitivity image of \vlbipsr observed by the \textit{Hubble Space Telescope} with the Wide-Field Planetary Camera 2 and the F814W filter (roughly I band) on 1996 May 19. We find an upper limit of  $m_\mathrm{I}=22$ for the optical counterparts of R1 and R2, in comparison with the apparent magnitude of \vlbipsr \citep[$m_\mathrm{F814W}=19.42$,][]{Durant2012}. 

It is quite difficult to find a radio star even in a dedicated survey. \citet{Kimball2009} searched for candidate radio stars by comparing the survey of the Faint Images of the Radio Sky at Twenty centimetre \citep[FIRST,][]{Becker1995} and the Sloan Digital Sky Survey (SDSS). They reported that for the million stars in the magnitude range $15<m_\mathrm{i}<19.2$, only $\leq$1.2 radio stars are detected with a flux of $S_\mathrm{1.4} \geq1.25$~mJy. Considering that the fraction of radio stars with flux densities above the FIRST limit declines at fainter optical magnitude,  the random probability of being a radio star is extremely small, $\la$1$\times$10$^{-6}$, for R1 or R2.   

Sources R1 and R2, interpreted as VLBI detected radio cores, are likely hosted in early type or bulge-dominated galaxies according to the statistical study of the faint sky by \citet{HerreraRuiz2017}. VLBI is a powerful tool for revealing AGNs in some early type galaxies without clear signs of nuclear activity in the optical and infrared bands \citep[e.g.,][]{Park2017}. Moreover, sources R1 and R2 likely belong to a group of infrared-faint radio sources (IFRS), which has a radio-to-infrared flux density ratio in the order of 100 or above \citep{Norris2006, Maini2016}.    

\subsection{Toward $\leq$10~$\umu$as VLBI astrometry}
\label{sec4-2}

Since sources R1 and R2 are most likely the compact cores in AGNs, they can be used as the in-beam phase-referencing sources to perform extremely high-precision astrometry on \vlbipsr. These are at least about 150 times closer to the pulsar than is the calibrator used by \citet{Deller2008}. Amongst the known in-beam phase referencing observations of pulsars \citep[e.g.,][]{Deller2013, Kirsten2015}, they are the closest reference sources. It is also the first time to have multiple reference sources each with a separation of less than one arcmin. While previous studies have rarely sought such faint reference sources (we can make use of fainter sources since \vlbipsr itself is bright enough to provide the calibration solutions), the presence of not one but two background sources so near a target is fortuitous. The VLA survey with an angular resolution of $\sim$5~arcsec \citep[FIRST, ][]{White1997} shows that the density of sources brighter than $\sim$1~mJy at 1.4~GHz is about 90 per square deg or one per 40 square arcmin. A simplistic calculation gives a $\sim$5 per cent chance of getting one background source, and just a $\sim$0.25 per cent chance of finding two, in a randomly selected area of one square arcmin. The chance that both are compact on mas scales is even lower again.  However, as shown above, we have convincing evidence that both sources are real and extragalactic in nature.

Differential VLBI astrometry on \vlbipsr with respect to the two extremely nearby sources will enable us to minimise potential systematic errors, caused by the phase variations arising from the propagation through the ionosphere and the troposphere, to a negligible level. Because the propagation-related components dominated the error budget, the earlier astrometric accuracy on \vlbipsr is limited to 54~$\umu$as \citep{Deller2008}. The systematic errors have a tight correlation with the separation between the calibrator and the target, $\sim$1--2~$\umu$as per arcmin \citep{Chatterjee2004, Deller2013, Kirsten2015}. According to this empirical dependence, systematic errors will not limit the astrometric accuracy until well below $\sim$10~$\umu$as, which is the current best achieved result \citep[e.g.,][]{Deller2013, Reid2014, Yang2016}. At this level, calibrator stability, rather than differential propagation errors, becomes dominant.

The two faint sources are likely extremely stable reference points. As the radio cores in extragalactic AGNs, they are the most stationary part of a jet structure in the stable accretion state. There is no evidence of a major radio outburst over the 19 yr in the past observations. Compared to bright radio sources, faint sources may have much more stable radio cores since they have either a much weaker Doppler beaming factor, a relatively higher redshift, or an intrinsically low accretion rate. With the multi-epoch European VLBI Network (EVN) in-beam phase-referencing observations at 5~GHz spanning three years, \citet{Yang2016} revealed that the radio core in a sub-mJy source has a positional stability of $\leq$13~$\umu$as per epoch, at least about one order of magnitude better than that of a bright source. Thus, it is quite promising for us to reach an accuracy of $\leq$10~$\umu$as before the core stability begins to dominate the error budget. The presence of the second in-beam calibrator provides a unique opportunity to directly constrain the potential contribution of the positional stability of the two radio cores on the astrometric accuracy. The advantages of multiple in-beam reference sources include calibration of time-varying phase errors on much shorter time-scales and higher resulting on-source durations for each source providing better SNRs, compared to multiple reference-source observations requiring antenna slewing to cross-check the positional stability across epochs \citep[e.g.,][]{Campbell1996} or to solve for a two-dimension propagation-induced phase-variation screen \citep[e.g.,][]{Rioja2017}. Over the next two years, we will make four astrometric observations of \vlbipsr with the more sensitive VLBI network including the new Chinese Tianma 65-m radio telescope \citep{An2017}, which will enable us to directly test these predictions.

The new VLBI astrometry on \vlbipsr can be performed at an observation frequency of $\la$6.7~GHz. The small angular separations amongst the three sources permit the inclusion of the largest radio telescope in the southern hemisphere, the Tidbinbilla 70-m radio telescope, to observe them within the FWHM of a single antenna beam at a frequency up to 22 GHz. However, \vlbipsr has a quite steep spectrum, $\propto\nu^{-2.0}_\mathrm{obs}$. To calibrate out the residual phase errors via the fringe-fitting and to self-calibrate on the pulsar data with a typical baseline sensitivity limit, 1$\sigma$ $\sim$2~mJy for two 25-m dishes, it is necessary to have a correlation amplitude of $\ga$10~mJy. Considering the spectral index of the pulsar, it is better to run the in-beam phase-referencing VLBI observations at a frequency of $\la$6.7~GHz. By applying gating during the correlation, the signal to noise ratio of the pulsar's fringes can be further improved by a factor between three from our 2.3 GHz observations and seven from the earlier 8.4 GHz observations \citep{Deller2008}. Moreover, both sources R1 and R2 will have slightly higher correlation amplitudes because of their partially optically thick spectra at $\la$6.7~GHz.   

Currently, the parallax accuracy of \vlbipsr can be improved by a factor of at least two with the in-beam phase-referencing observations. The final VLBI image with naturally weighting will have a beam size of $\sim$1~mas at 6.7~GHz with the currently available VLBI capabilities. With the available bandwidth of the LBA (i.e. 64~MHz), an on-source time of 300~min, and all these telescopes, one can get a practical image sensitivity of 1$\sigma$ $\sim$0.05~mJy~beam$^{-1}$. Both sources R1 and R2 have a flux density of $\sim$1~mJy at 6.7~GHz. Thus, the VLBI images will have a SNR of 20 and a statistical astrometric accuracy of $\frac{\theta_\mathrm{b}}{2\mathrm{SNR}}\sim25$~$\umu$as for each source.  Even with only three-epoch VLBI observations, a parallax accuracy of $\la$25~$\umu$as is also achievable. In the future, the parallax accuracy of \vlbipsr can be continuously enhanced toward to an extremely high level, $\leq$10~$\umu$as with the wide-band VLBI backends \citep[e.g., Digital Base-Band Converters 3,][]{Tuccari2014} and more VLBI stations: the Warkworth 30-m radio telescope \citep{Petrov2015, Woodburn2015}, the Nkutunse 32-m radio telescope at Kuntunse \citep[e.g.,][]{Duah2014}, and the SKA-Mid \citep{Paragi2015}. 

The finding of the two sources will lead to not only a more accurate $D_\mathrm{\pi}$ for \vlbipsr but also a more stringent constraint on $\dot{G}$ in the near future. The relationship between $D_\mathrm{\pi}$ and $\dot{G}$ can be simply written as \citep[e.g.,][]{Deller2008}:
\begin{equation} 
\frac{\dot G}{G}  =-\frac{\mu^2}{2c}(D_\mathrm{k}-D_\mathrm{\pi}),
 \label{eq2}
\end{equation} 
where $c$ is the light speed and $\mu$ is the proper motion. Compared to the VLBI distance of $D_\mathrm{\pi}$, the pulsar timing distance of $D_\mathrm{k}$ in Equation \ref{eq2} includes some excess contributions from changes intrinsic to the pulsar system, e.g., $\dot{G}$, and the differential acceleration of the Solar system and the pulsar system. Currently, the $\mu$ has been measured with quite high fractional accuracy \citep[$2\times10^{-5}$,][]{Reardon2016}. The $D_\mathrm{\pi}$ has a fractional accuracy of 0.83 per cent \citep{Deller2008}, about five times worse than the latest measurement of $D_\mathrm{k}$ \citep[0.16 per cent,][]{Reardon2016}. Improvements to $D_\mathrm{\pi}$ will therefore have a direct and marked effect on the constraint on $\dot{G}/G$. Since the accuracy of $D_\mathrm{k}$ will rapidly improve further as the uncertainty on the observed orbital period derivative decreases with ongoing timing observations, a future comparison in which both $D_\mathrm{k}$ and $D_\mathrm{\pi}$ are measured to better than one part in 1000 is foreseeable, which would lead to an order-of-magnitude improvement in the uncertainty of $\dot{G}/G$.

\section{Conclusions}
\label{sec5}
To achieve a VLBI parallax distance with a sub-pc accuracy for \vlbipsr, the nearest and brightest radio millisecond pulsar, via in-beam phase-referencing observations, we searched for compact radio sources near the pulsar. Within a circle of one-arcmin radius, we detected two mJy radio sources using publicly available VLA and ATCA data. We also carried out VLBI observations of the two sources with the LBA plus the Kunming radio telescope at 2.3~GHz. Their VLBI images show that each has a compact structure and a brightness temperature of $\geq$10$^7$~K. They have a flat spectrum and a relatively stable flux density over 19~yr. In the existing deep optical and X-ray images of \vlbipsr, we found no counterparts. Combining all these results, we argue that they are most likely compact radio cores in AGNs rather than Galactic radio stars. The existence of the two extremely nearby radio sources makes it quite promising to achieve a VLBI distance measurement at a sub-pc accuracy with the current VLBI network configuration and at a sub-light-year accuracy when more stations and wider bandwidth are available. These continuous improvements will also directly lead to a more stringent constraint on the time stability of Newton's gravitational constant $G$.  

\section*{Acknowledgements}
This work was partly supported by the SKA pre-construction funding from the Ministry of Science and Technology of China (2016YFE0100300) and the Chinese Academy of Sciences (CAS).  TA thanks the grant supported by the Youth Innovation Promotion Association of CAS and FAST Fellowship which is supported by the Centre for Astronomical Mega-Science, CAS. We thank  Chris Phillips (the Australia Telescope National Facility), Zhiqiang Shen (Shanghai Astrnomical Observatory), and Min Wang (Yunnan Astronomical Observatories) for scheduling their telescope time, and the station operators and engineers for supporting the joint VLBI experiment. JY thanks J. Quick, L.I. Gurvits, and G. Hobbs for the helpful discussion. The Australia Telescope Compact Array, the Parkes radio telescope, the Mopra radio telescope and the Long Baseline Array are part of the Australia Telescope National Facility which is funded by the Australian Government for operation as a National Facility managed by CSIRO. This work made use of the Swinburne University of Technology software correlator, developed as part of the Australian Major National Research Facilities Programme. This paper includes archived data obtained through the Australia Telescope Online Archive (http://atoa.atnf.csiro.au). The National Radio Astronomy Observatory is a facility of the National Science Foundation operated under cooperative agreement by Associated Universities, Inc.





\begin{thebibliography}{99}

\bibitem[\protect\citeauthoryear{Alpar et al.}{1982}]{Alpar1982}
Alpar M.~A., Cheng A.~F., Ruderman M.~A., Shaham J., 1982, \nat, 300, 728 

\bibitem[\protect\citeauthoryear{An, Sohn \& Imai}{2017}]{An2017}
An T., Sohn B.-W., Imai H., 2017, Nat. Astron., in press

\bibitem[\protect\citeauthoryear{Becker, White \& Helfand}{1995}]{Becker1995}
Becker R.~H., White R.~L., Helfand D.~J., 1995, \apj, 450, 559

\bibitem[\protect\citeauthoryear{Bogdanov}{2013}]{Bogdanov2013}
Bogdanov S., 2013, \apj, 762, 96

\bibitem[\protect\citeauthoryear{Campbell et al.}{1996}]{Campbell1996}
Campbell R.~M., Bartel N., Shapiro I.~I., Ratner M.~I., Cappallo R.~J., Whitney A.~R., Putnam N., 1996, \apj, 461, L95

\bibitem[\protect\citeauthoryear{Chatterjee et al.}{2004}]{Chatterjee2004}
Chatterjee S., Cordes J.~M., Vlemmings W.~H., Arzoumanian Z., Goss W.~M., Lazio T.~J., 2004, \apj, 604, 339

\bibitem[\protect\citeauthoryear{Condon et al.}{1982}]{Condon1982}
Condon J.~J., Condon M.~A., Gisler G., Puschell J.~J., 1982, \apj, 252, 102

\bibitem[\protect\citeauthoryear{Dai et al.}{2015}]{Dai2015}
Dai S. et al., 2015, \mnras, 449, 3223

\bibitem[\protect\citeauthoryear{Deller et al.}{2007}]{Deller2007}
Deller A.~T., Tingay S.~J., Bailes M., West C., 2007, \pasp, 119, 318

\bibitem[\protect\citeauthoryear{Deller et al.}{2008}]{Deller2008}
Deller A.~T., Verbiest J.~P.~W., Tingay S.~J., Bailes M., 2008, \apj, 685, L67

\bibitem[\protect\citeauthoryear{Deller et al.}{2013}]{Deller2013}
Deller A.~T., Boyles J., Lorimer D.~R., Kaspi V.~M., McLaughlin M.~A., Ransom S., Stairs I.~H., Stovall K., 2013, \apj, 770, 145 

\bibitem[\protect\citeauthoryear{Deller et al.}{2016}]{Deller2016}
Deller A.~T. et al., 2016, \apj, 828, 8

\bibitem[\protect\citeauthoryear{Duah Asabere et al.}{2014}]{Duah2014}
Duah Asabere B., Gaylard M., Horellou C., Winkler H., Jarrett T., 2014, in Engelbrecht C., Karataglidis S., eds., Proceedings of SAIP2014, the 59th Annual Conference of the South African Institute of Physics, p. 296, ISBN: 978-0-620-65391-6 

\bibitem[\protect\citeauthoryear{Durant et al.}{2012}]{Durant2012}
Durant M. et al., 2012, \apj, 746, 6 

\bibitem[\protect\citeauthoryear{Greisen}{2003}]{Greisen2003} 
Greisen E.~W., 2003, in Heck A., eds., Astrophysics and Space Science Library, Vol. 285, Information Handling in Astronomy: Historical Vistas. Kluwer, Dordrecht, p. 109

\bibitem[\protect\citeauthoryear{Hao, Wang \& Yang}{2010}]{Hao2010} 
Hao L.-F., Wang M., Yang J., 2010, Res. Astron. Astrophys., 10, 805

\bibitem[\protect\citeauthoryear{Herrera Ruiz et al.}{2017}]{HerreraRuiz2017} 
Herrera Ruiz N., et al., 2017, \aap, 607, A132

\bibitem[\protect\citeauthoryear{Johnston et al.}{1993}]{Johnston1993} 
Johnston S. et al., 1993, \nat, 361, 613

\bibitem[\protect\citeauthoryear{Kimball et al.}{2008}]{Kimball2008} 
Kimball A.~E., Ivezi\'c \v{Z}., 2008, \aj, 136, 684

\bibitem[\protect\citeauthoryear{Kimball et al.}{2009}]{Kimball2009} 
Kimball A.~E., Knapp G.~R., Ivezi\'c \v{Z}., West A.~A., Bochanski J.~J., Plotkin R.~M., Gordon, M.~S., 2009, \apj, 701, 535

\bibitem[\protect\citeauthoryear{Kirsten et al.}{2015}]{Kirsten2015} 
Kirsten F., Vlemmings W., Campbell R.~M., Kramer M., Chatterjee S., 2015, \aap, 577, 111

\bibitem[\protect\citeauthoryear{Maini et al.}{2016}]{Maini2016} 
Maini A., Prandoni~I., Norris R.~P., Spitler L.~R., Mignano A., Lacy M., Morganti R., 2016, \aap, 596, A80 

\bibitem[\protect\citeauthoryear{Manchester}{2015}]{Manchester2015} 
Manchester R.~N., 2015, International Journal of Modern Physics D, 24, id. 1530018

\bibitem[\protect\citeauthoryear{Norris et al.}{2006}]{Norris2006} 
Norris R.~P. et al., 2006, \aj, 132, 2409

\bibitem[\protect\citeauthoryear{Paragi et al.}{2015}]{Paragi2015} 
Paragi Z. et al., 2015, Proceedings of Advancing Astrophysics with the Square Kilometre Array, PoS (AASKA14), 143

\bibitem[\protect\citeauthoryear{Park et al.}{2017}]{Park2017} 
Park S., Yang J., Oonk J.~B.~R., Paragi Z., 2017, \mnras, 465, 3943

\bibitem[\protect\citeauthoryear{Petrov et al.}{2015}]{Petrov2015} 
Petrov L., Natusch T., Weston S., McCallum J., Ellingsen S., Gulyaev S.,  2015, \pasp, 127, 516

\bibitem[\protect\citeauthoryear{Reardon et al.}{2016}]{Reardon2016} 
Reardon D.~J. et al., 2016, \mnras, 455, 1751

\bibitem[\protect\citeauthoryear{Reid \& Honma}{2014}]{Reid2014} 
Reid M.~J., Honma, M., 2014, \araa, 52, 339

\bibitem[\protect\citeauthoryear{Rioja et al.}{2017}]{Rioja2017} 
Rioja M.~J., Dodson R., Orosz G., Imai H., Frey S., 2017, \aj, 153, 105

\bibitem[\protect\citeauthoryear{Sault, Teuber \& Wright}{1995}]{Sault1995} 
Sault R.~J., Teuben P.~J., Wright M.~C.~H., 1995, in Shaw R.~A., Payne H.~E., Hayes J.~J.~E., eds., ASP Conf. Ser., Vol. 77, Astronomical data Analysis Software and Systems IV, p. 433

\bibitem[\protect\citeauthoryear{Shannon et al.}{2015}]{Shannon2015} 
Shannon R.~M. et al., 2015, \sci, 349, 1522

\bibitem[\protect\citeauthoryear{Shepherd, Pearson \& Taylor}{1994}]{Shepherd1994} 
Shepherd M.~C., Pearson T.~J., Taylor G.~B., 1994, \baas, 26, 987

\bibitem[\protect\citeauthoryear{Tingay et al.}{2003}]{Tingay2003} 
Tingay S.~J. et al., 2003, \pasj, 55, 351

\bibitem[\protect\citeauthoryear{Tuccari et al.}{2014}]{Tuccari2014} 
Tuccari G., Walter A., Salvatore B., Casey S., Felke A., Lindqvist M.,  2014, in Behrend D., Baver K.~D.,  Armstrong K. L., eds., International VLBI Service for Geodesy and Astrometry 2014 General Meeting Proceedings: "VGOS: The New VLBI Network", p. 86, ISBN 978-7-03-042974-2

\bibitem[\protect\citeauthoryear{Verbiest et al.}{2008}]{Verbiest2008} 
Verbiest J.~P.~W. et al., 2008, \apj, 679, 675

\bibitem[\protect\citeauthoryear{White et al.}{1997}]{White1997} 
White R.~L., Becker R.~H., Helfand D.~J.,  Gregg M.~D.,  1997, \apj, 475, 479

\bibitem[\protect\citeauthoryear{Woodburn et al.}{2015}]{Woodburn2015} 
Woodburn L., Natusch T., Weston S., Thomasson P., Godwin M., Granet C., Gulyaev S.,  2015, \pasa, 32, 17

\bibitem[\protect\citeauthoryear{Yang et al.}{2016}]{Yang2016} 
Yang J., Paragi Z., van der Horst A.~J., Gurvits L.~I., Campbell R.~M., Giannios~D., An T., Komossa,  2016, \mnras, 462, L66

\bibitem[\protect\citeauthoryear{Zavlin et al.}{2002}]{Zavlin2002} 
Zavlin V.~E., Pavlov G.~G., Sanwal D., Manchester R.~N., Tr\"{u}mper J., Halpern J. ~P., Becker W.,  2002, \apj, 569, 894


\end{thebibliography}

%



\appendix




\bsp	
\label{lastpage}
\end{document}